\documentclass[
reprint,
 amsmath,amssymb,
 aps,
 pra,
superscriptaddress
]{revtex4-2}
\usepackage{siunitx}
\usepackage{graphicx}
\usepackage{dcolumn}
\usepackage{bm}
\usepackage{subcaption}
\usepackage{tikz}
\usetikzlibrary{calc}
\usetikzlibrary{arrows.meta,arrows}
\DeclareSIUnit\gauss{G}
\DeclareSIUnit\ppm{ppm}
\DeclareSIUnit\phonon{phonon}
\newcommand{\ket}[1]{|#1\rangle}

\newcommand{\sidecaption}[1]
{\raisebox{\abovecaptionskip}{\begin{subfigure}[t]{2em}
  \caption[singlelinecheck=off]{}
  \label{#1}
\end{subfigure}}\ignorespaces}

\usepackage{comment}
\begin{document}


\title{Integrating a fiber cavity into a wheel trap for strong ion-cavity coupling}

\author{Markus Teller}
\affiliation{Institut für Experimentalphysik, Universität Innsbruck, Technikerstrasse 25, 6020 Innsbruck, Austria}
\author{Viktor Messerer}
\affiliation{Institut für Experimentalphysik, Universität Innsbruck, Technikerstrasse 25, 6020 Innsbruck, Austria}
\author{Klemens Sch\"{u}ppert}
\altaffiliation{Present address: Infineon Technologies Austria AG, Siemensstraße 2, 9500, Villach, Austria}
\affiliation{Institut für Experimentalphysik, Universität Innsbruck, Technikerstrasse 25, 6020 Innsbruck, Austria}
\author{Yueyang Zou}
\affiliation{Institut für Experimentalphysik, Universität Innsbruck, Technikerstrasse 25, 6020 Innsbruck, Austria}
\author{Dario A. Fioretto}
\affiliation{Institut für Experimentalphysik, Universität Innsbruck, Technikerstrasse 25, 6020 Innsbruck, Austria}
\author{Maria Galli}
\affiliation{Institut für Experimentalphysik, Universität Innsbruck, Technikerstrasse 25, 6020 Innsbruck, Austria}
\author{Philip C. Holz}
\affiliation{Institut für Experimentalphysik, Universität Innsbruck, Technikerstrasse 25, 6020 Innsbruck, Austria}
\affiliation{Alpine Quantum Technologies GmbH, Technikerstrasse 17/1, 6020 Innsbruck, Austria}
\author{Jakob Reichel}
\affiliation{Laboratoire Kastler Brossel, ENS-Universit\'e PSL, CNRS, Sorbonne Universit\'e, Coll\`ege de France 24 rue Lhomond, 75005 Paris, France}
\author{Tracy E. Northup}
\affiliation{Institut für Experimentalphysik, Universität Innsbruck, Technikerstrasse 25, 6020 Innsbruck, Austria}
\date{\today}

\begin{abstract}
We present an ion trap with an integrated fiber cavity, designed for strong coupling at the level of single ions and photons.  The cavity is aligned to the axis of a miniature linear Paul trap, enabling simultaneous coupling of multiple ions to the cavity field. We simulate how charges on the fiber mirrors affect the trap potential, and we test these predictions with an ion trapped in the cavity.  Furthermore, we measure micromotion and heating rates in the setup.
\end{abstract}
\maketitle

\section{\label{sec:Intro}Introduction}

In a quantum network, a coherent interface consists of a unitary interaction between light and matter that is much stronger than decay channels to the environment~\cite{Kimble2008a}.
Recent experiments with a single trapped ion coupled to a fiber-based optical resonator have demonstrated a coherent coupling rate $g_0$ exceeding the atomic spontaneous-emission rate $\gamma$~\cite{Steiner2013,Ballance2017,Takahashi2020,Christoforou2020}. 
Due to this coherent ion--photon interaction, fiber-based cavities integrated with ion traps offer a promising platform for a quantum network node.
Additional strengths of the platform include both direct emission into optical fibers, for transmission in long-distance networks, and the small fiber footprint, for miniaturizing quantum nodes and scaling up their complexity.

Many key applications of quantum networks, including distributed quantum computation, require the network nodes to host more than a single qubit~\cite{Wehner2018,cuomo2020towards}.
While a single trapped ion has been coupled to a fiber cavity~\cite{Steiner2013,Kassa2018}, 
coupling of multiple ions has not yet been achieved.
The ion-trap geometries used in Refs.~\cite{Steiner2013,Kassa2018} give rise to residual radiofrequency fields that vanish only at a single point within the trapping region and introduce excess micromotion away from this point. It is not possible to compensate for this micromotion~\cite{Berkeland1998}, which compromises both the ion--cavity coupling~\cite{Kassa2018} and the fidelity of gate operations between multiple ions.

Here, we present an ion--cavity system designed for strong coupling of multiple ions to a fiber cavity. The fiber mirrors are integrated along the axis of a linear Paul trap known as a wheel trap.  Ions can be positioned along this axis without introducing excess micromotion.
The paper is structured as follows: Section~\ref{sec:Trap} describes the ion-cavity system. The influence of surface charges and the influence of the fiber-mirror positions on the ion are studied with simulations in Sec.~\ref{sec:Trap_simulation}. We then turn to experiments in Sec.~\ref{sec:Measurements}, first with a test setup without integrated fibers, and subsequently with the integrated ion--cavity system. We present measurements of micromotion in the test setup and heating rates of both the test setup and the ion--cavity system. As a final test, we show that the effects of surface charges, commonly found on dielectric surfaces~\cite{Harlander2010,Ong2020probing}, can be counteracted by means of the trap electrodes.

\section{\label{sec:Trap}Experimental setup}

We first focus on the details of the wheel trap, a linear Paul trap developed for quantum metrology experiments~\cite{Chen2017,chen2017sympathetic,Brewer2019}. The trap consists of a diamond wafer, $\SI{300}{\micro\meter}$ thick, on which gold electrodes are sputtered,  $\SI{5}{\micro\meter}$ thick, using titanium as an adhesive layer.
The thinness of this wafer makes the wheel trap uniquely suited for integration with an optical microcavity along the trap axis.

A photo of our adaptation of the wheel trap is shown in Fig.~\ref{fig:trap1}, and a schematic of the trap center is shown in Fig.~\ref{fig:trap2}.  We define the x-y plane as the plane of the wafer and the z axis as the trap axis. The ion--electrode distance is $\SI{250}{\micro\meter}$. 
It is important to match the capacitances of the four RF electrodes. In the trap-design phase, we calculate these capacitances using finite-element-analysis software and adjust the electrode geometry accordingly. Compensation electrodes are used to minimize micromotion in the x-y plane~\cite{Berkeland1998,Keller2015}. In contrast to earlier wheel-trap designs~\cite{Chen2017,chen2017sympathetic,Brewer2019}, the DC electrodes (also known as endcaps) that confine ions along the z axis are hollow, allowing us to integrate fiber mirrors within them. The electrodes consist of stainless-steel tubes with an inner diameter of $\SI{250(50)}{\micro\meter}$ and an outer diameter of $\SI{500(20)}{\micro\meter}$. 

\begin{figure}
\begin{subfigure}{0.38\textwidth}
\sidecaption{fig:trap1}
\raisebox{-\height}{
\begin{tikzpicture}
\node [xscale=1,
    above right,
    inner sep=0] (image) at (0,0) {\includegraphics[scale=0.2]{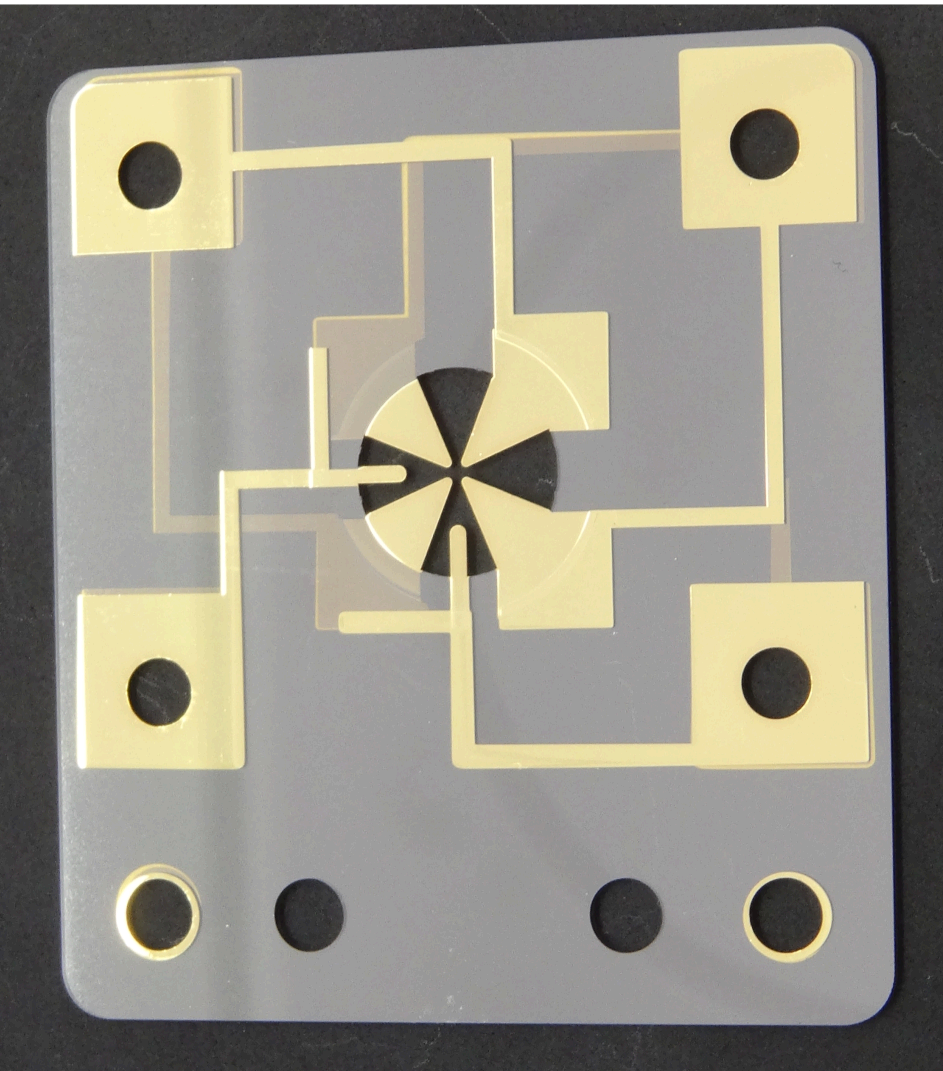}};
		\begin{scope}[
x={($0.1*(image.south east)$)},
y={($0.1*(image.north west)$)}]
		\draw[{Stealth[length=4mm, width=2.7mm]}-,very thick] (4.2,5.55) -- (2.4,10.15) node[above,black,fill=white,align=center]{\small Compensation\\ electrode};
		\draw[{Stealth[length=4mm, width=2.7mm]}-,very thick] (5.5,6.2) -- (7.7,10.15) node[above,black,fill=white,align=center]{\small RF electrode};
		\draw[{Stealth[length=4mm, width=2.7mm]}-,very thick] (0.65,2.1) edge (5,2.1) (9.45,2.1) -- (5,2.1) node[above,black,fill=none,align=center]{\small $\SI{25}{\milli\meter}$};
		\draw[{Stealth[length=4mm, width=2.7mm]}-,very thick] (4,1) -- (3,-0.15) node[below,black,fill=white,align=center]{\small Diamond wafer};
\end{scope}
\end{tikzpicture}}
\end{subfigure}
\begin{subfigure}{0.41\textwidth}
\sidecaption{fig:trap2}
\raisebox{-\height}{
\begin{tikzpicture}
\node [xscale=1,
    above right,
    inner sep=0] (image) at (0,0) {\includegraphics[scale=0.2]{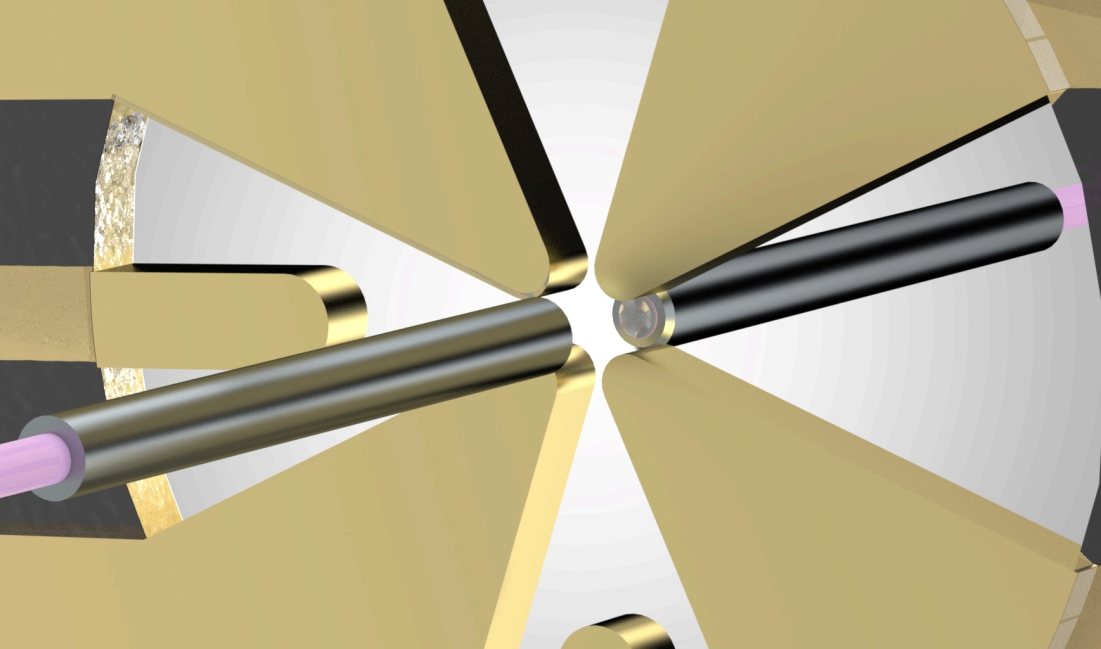}};
		\begin{scope}[
x={($0.1*(image.south east)$)},
y={($0.1*(image.north west)$)}]
		\draw[{Stealth[length=4mm, width=2.7mm]}-,very thick] (0.35,3) -- (1.2,10.15)node[above,black,fill=white,align=center]{\small Fiber};
		\draw[{Stealth[length=4mm, width=2.7mm]}-,very thick] (3.5,8) edge (8,10.15) (6.6,8) -- (8,10.15) node[above,black,fill=white,align=center]{\small RF electrodes};
		
		\draw[{Stealth[length=4mm, width=2.7mm]}-,very thick] (3.5,4) edge (8,-0.15) (7,5.5) -- (8,-0.15) node[below,black,fill=white,align=center]{\small DC electrodes\\ with fiber mirrors};
		
		\draw[{Stealth[length=4mm, width=2.7mm]}-,very thick] (2,5) -- (2.5,-0.15);
		\draw[{Stealth[length=4mm, width=2.7mm]}-,very thick] (5.5,0) -- (4,-0.5) node[anchor=north east,yshift=2mm,black,fill=none,align=center]{\small Compensation\\ electrodes};
		\draw[-,very thick] (0.1,0.5) edge (0.9,0.5) (1.95,0.5) -- (0.9,0.5) node[above,black,fill=none,align=center]{\small $\SI{1}{\milli\meter}$};
		\draw[-{Stealth},very thick] (8.8,4) -- ++ (0.6,-1.29) node[below,black,fill=none,align=center,yshift=1mm]{\small x};
		\draw[-{Stealth},very thick] (8.8,4) -- ++ (-0.8,-.4) node[left,black,fill=none,align=center,xshift=1mm]{\small z};
		\draw[-{Stealth},very thick] (8.8,4) -- ++ (0.65,1.2) node[above,black,fill=none,align=center,yshift=-1mm]{\small y};
\end{scope}
\end{tikzpicture}}
\end{subfigure}
\caption{a) Image of a wheel trap. b) Schematic of the ion-cavity system.}
\end{figure}
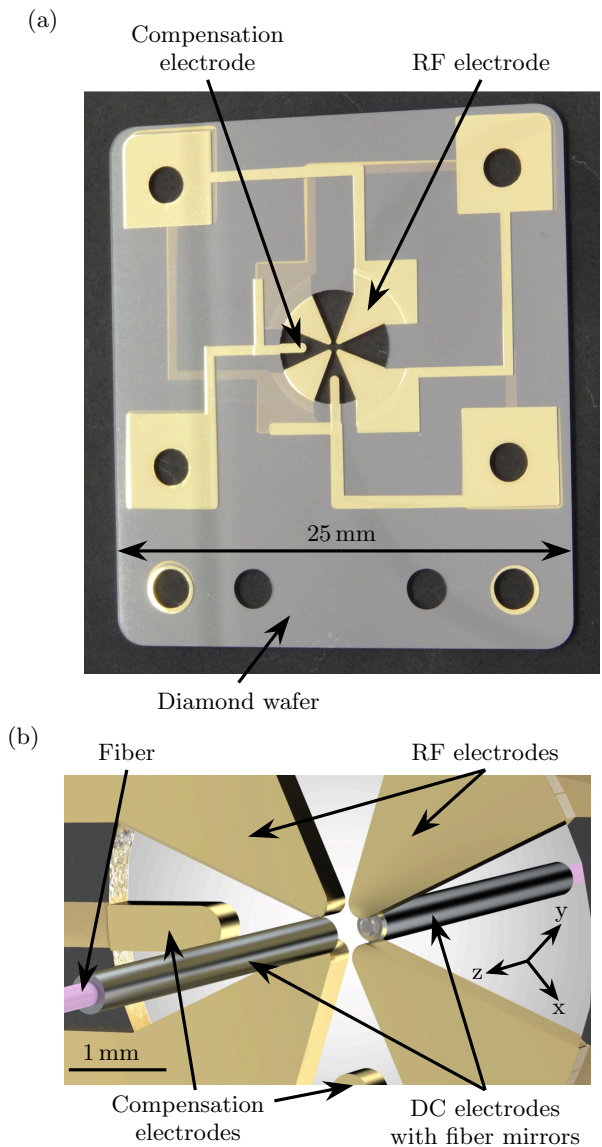

The two fiber mirrors inside the opposing endcaps form a fiber Fabry-Pérot cavity (FFPC)~\cite{Hunger2010}.
To prevent charging of the fibers due to scattered laser light~\cite{Harlander2010,Ong2020probing}, we recess the fibers by $\SI{10(2)}{\micro\meter}$ into the tubes.
We use one multimode (MM) fiber~\footnote{IVG-Fiber Cu200/220} with a diameter of $\SI{220(3)}{\micro\meter}$ and one photonic-crystal (PC) fiber~\footnote{NKT Photonics PCF - LMA 20} with a diameter of $\SI{230(5)}{\micro\meter}$. A CO$_2$ laser-ablation process generates concave, near-spherical profiles on each fiber facet~\cite{Hunger2010,Ott2016a}. The radii of curvature are $\SI{318(5)}{\micro\meter}$ for the MM profile and $\SI{312(5)}{\micro\meter}$ for the PC profile. 
The mirrors consist of alternating layers of SiO$_2$ and Ta$_2$O$_5$, applied to the fiber facets via ion-beam sputtering~\footnote{Advanced Thin Films, Boulder, CO 80301, USA}.

At a wavelength of $\SI{854}{\nano\meter}$, the MM mirror has a transmission of $\SI{2(1)}{\ppm}$, whereas the transmission of the PC mirror is $\SI{16(1)}{\ppm}$. 
The cavity finesse is $\SI{9.2(2)e4}{}$ for a length of $\SI{507(8)}{\micro\meter}$, corresponding to a linewidth of $\kappa = 2\pi\cdot \SI{1.61(3)}{\mega\hertz}$ (half-width at half maximum). We calculate an ion-cavity coupling strength of $g_0 = 2\pi\cdot \SI{20.3(3)}{\mega\hertz}$ for the $\ket{3^2\mathrm{D}_{5/2}}$ to $\ket{4^2\mathrm{P}_{3/2}}$ transition of a $^{40}$Ca$^+$ ion~\cite{kimble1998strong}. 
The largest Clebsch-Gordon coefficient for transitions between Zeeman states of these manifolds is $\alpha = \sqrt{2/3}$, so that the largest possible coupling strength is $g = \alpha g_0 = 2\pi\cdot \SI{16.6(3)}{\mega\hertz}$.
The two relevant spontaneous emission channels are from $\ket{4^2\mathrm{P}_{3/2}}$ to $\ket{3^2\mathrm{D}_{5/2}}$, with a decay rate of $\gamma_\textrm{PD} = 2\pi\cdot \SI{0.67}{\mega\hertz}$, and from $\ket{4^2\mathrm{P}_{3/2}}$ to $\ket{4^2\mathrm{S}_{1/2}}$, with a decay rate of $\gamma_\textrm{PS} = 2\pi\cdot \SI{10.74}{\mega\hertz}$; both rates are half widths.
Based on the values of $g$, $\kappa$, $\gamma_\textrm{PD}$, and $\gamma_\textrm{PS}$, we expect our system to operate in the strong coupling regime~\cite{kimble1998strong}.

\begin{figure*}
\begin{subfigure}{1\textwidth}
\sidecaption{fig:figure2a}
\raisebox{-\height}{
\begin{tikzpicture}
\node [xscale=1,above right,inner sep=0] (image) at (0,0) {\includegraphics[width = 0.76\textwidth]{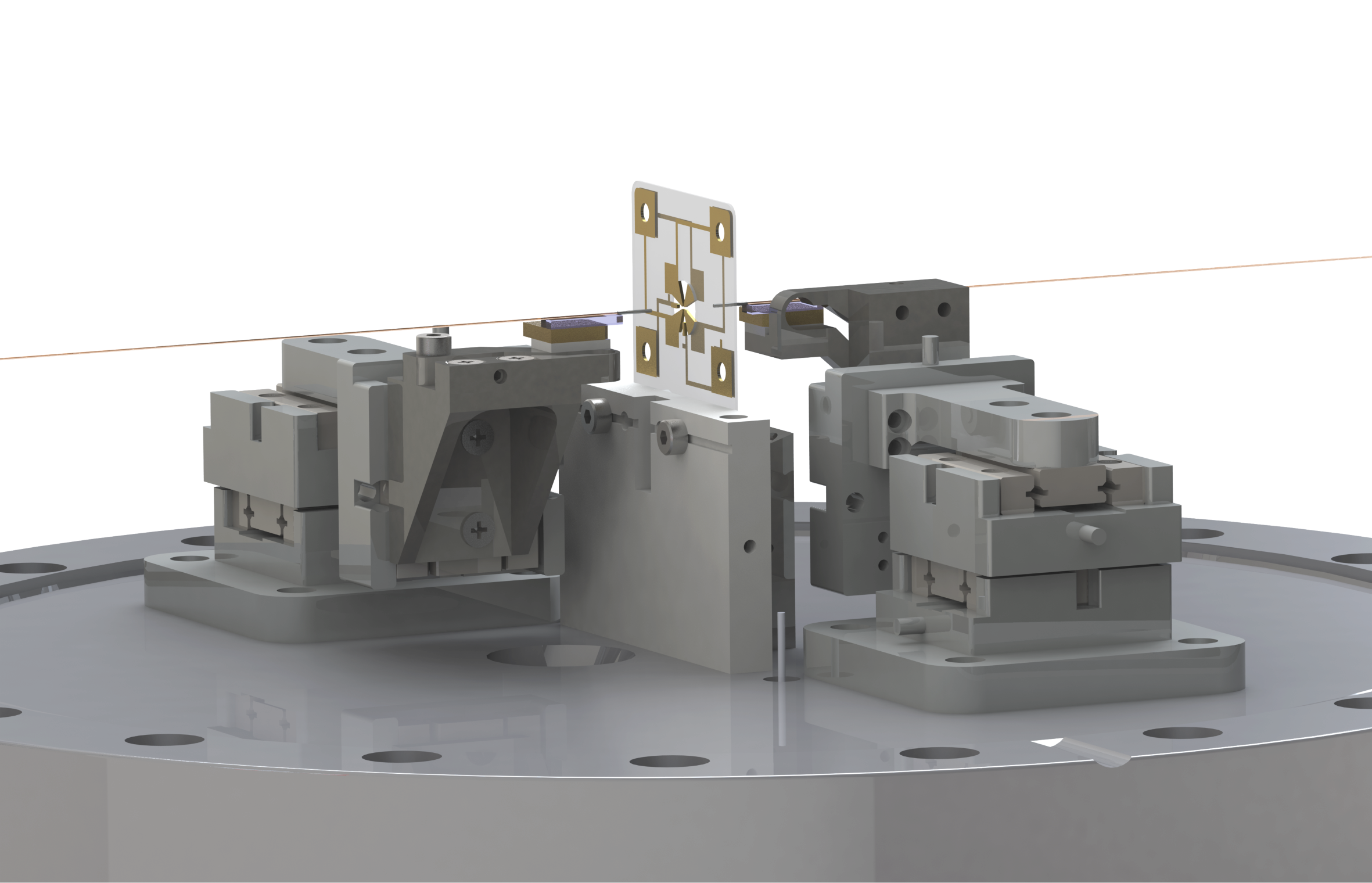}};
		\begin{scope}[
x={($0.1*(image.south east)$)},
y={($0.1*(image.north west)$)}]
		\node [xscale=1,
    above right,
    inner sep=0, anchor = north west,draw] (inset) at (0,10) {\includegraphics[width = 0.31\textwidth]{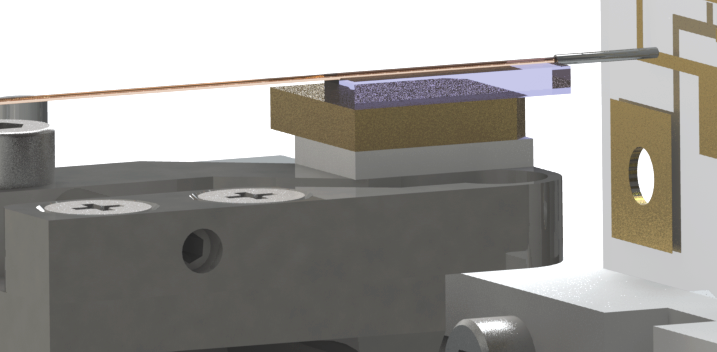}};
		\draw[{Stealth[length=4mm, width=2.7mm]}-,very thick] (7,6.8) -- (8,8.3) node[right,black,fill=none,align=center]{\small Vertical\\ tilt-adjuster};
		\draw[{Stealth[length=4mm, width=2.7mm]}-,very thick] (7.65,5.5) -- (8.8,6.2) node[above,black,fill=none,align=center]{\small 3D nanopositioning\\ assembly};
		\draw[dotted,very thick] ($(inset.south west)$) -- (3.2,5.5);
		\draw[dotted,very thick] ($(inset.south east)$) -- (4.8,6.5);
\end{scope}
		\begin{scope}[
		shift = ($(inset.south west)$),
x={($0.1*(inset.south east)$)},
y={($0.1*(inset.north west)$)}]
		\draw[{Stealth[length=4mm, width=2.7mm]}-,very thick] (9.2,8.6) -- (10.5,10) node[right,black,fill=none,align=center]{\small DC electrode with fiber mirror};
		\draw[{Stealth[length=4mm, width=2.7mm]}-,very thick] (8,7.7) -- (10.5,8.25) node[right,black,fill=none,align=center]{\small Quartz v-groove};
		\draw[{Stealth[length=4mm, width=2.7mm]}-,very thick] (7.3,6.5) -- (10.5,6.5) node[right,black,fill=none,align=center]{\small Shear-mode piezo};
		\draw[{Stealth[length=4mm, width=2.7mm]}-,very thick] (7.8,4) -- (10.5,4.75) node[right,black,fill=none,align=center]{\small Horizontal tilt-adjuster};
\end{scope}
\end{tikzpicture}}
\end{subfigure}
\begin{subfigure}{1\textwidth}
\sidecaption{fig:figure2b}
\raisebox{-\height}{
\begin{tikzpicture}
\node [xscale=1,
    above right,
    inner sep=0] (image) at (0,0) {\includegraphics[trim=0 190 0 85,clip,width = 
0.76\textwidth]{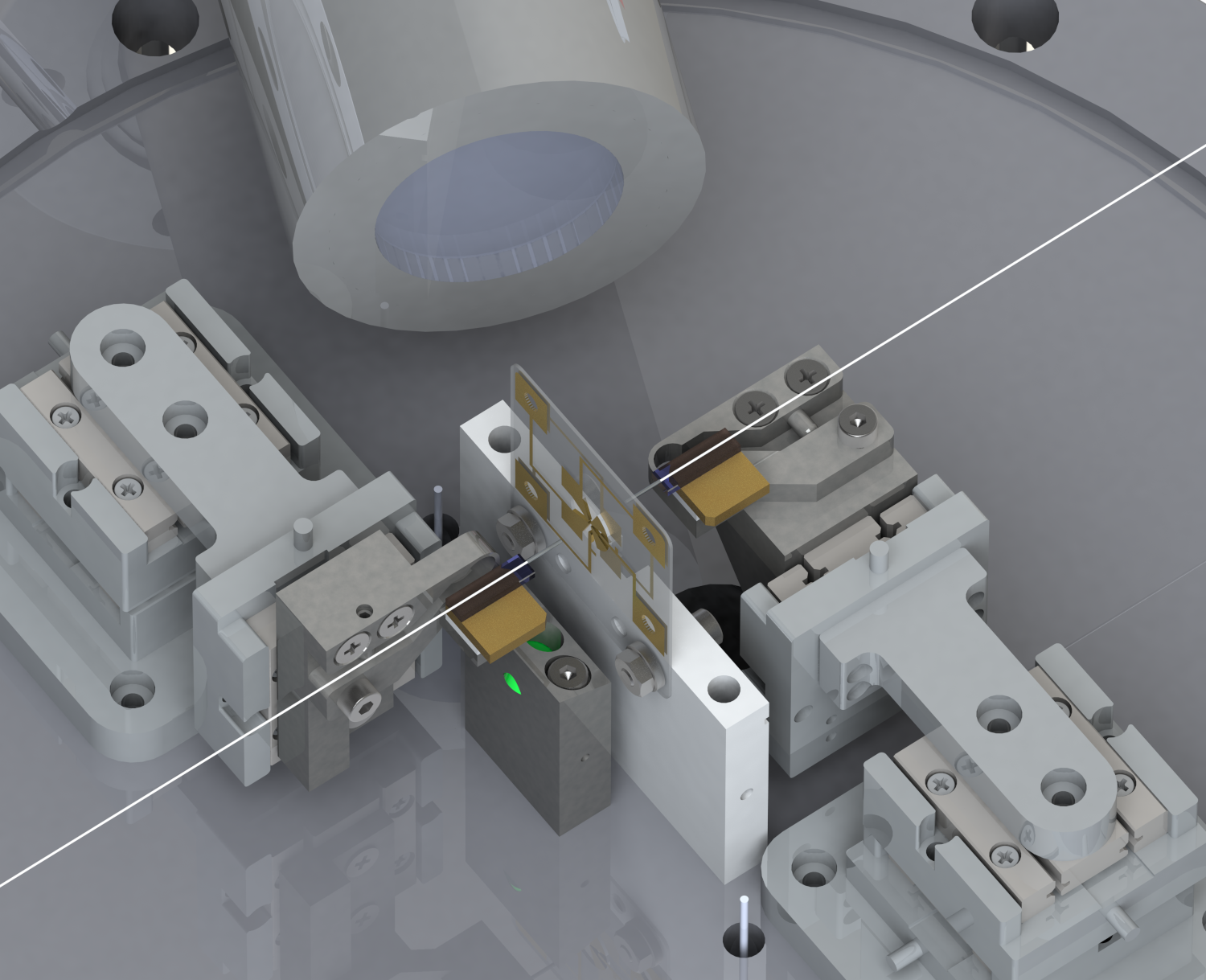}};
		\begin{scope}[
x={($0.1*(image.south east)$)},
y={($0.1*(image.north west)$)}]
		\draw[{Stealth[length=4mm, width=2.7mm]}-,very thick] (3.5,9.5) -- (3,10.15)node[above,black,fill=none,align=center]{\small Inverted viewport};
		\draw[{Stealth[length=4mm, width=2.7mm]}-,very thick] (6.5,6) -- (7,10.15)node[above,black,fill=none,align=center]{\small Horizontal\\ tilt-adjuster};
		
		\draw[{Stealth[length=4mm, width=2.7mm]}-,very thick] (2,1) -- (1.8,-0.15) node[below,black,fill=none,align=center]{\small 3D nanopositioning\\ assembly};
		\draw[{Stealth[length=4mm, width=2.7mm]}-,very thick] (3.3,1.5) -- (3.8,-0.15) node[below,black,fill=none,align=center]{\small Vertical\\ tilt-adjuster};
		\draw[{Stealth[length=4mm, width=2.7mm]}-,very thick] (4.46,1.85) -- (5.5,-0.15) node[below,black,fill=none,align=center]{\small Ablation target};
\end{scope}
\end{tikzpicture}}
\end{subfigure}
\caption{Rendered image of the ion-cavity system. a) Side view: To the left and right of the wheel trap, we see the DC electrodes with fiber mirrors, the quartz v-grooves, the shear-mode piezos, and the tilt-adjusters, all mounted on two 3D nanopositioning assemblies. b) Top view: An ablation target, along the line of sight of the wheel trap, is indicated in green. In this image, one of two inverted viewports is shown.}\label{fig:figure2}
\end{figure*}
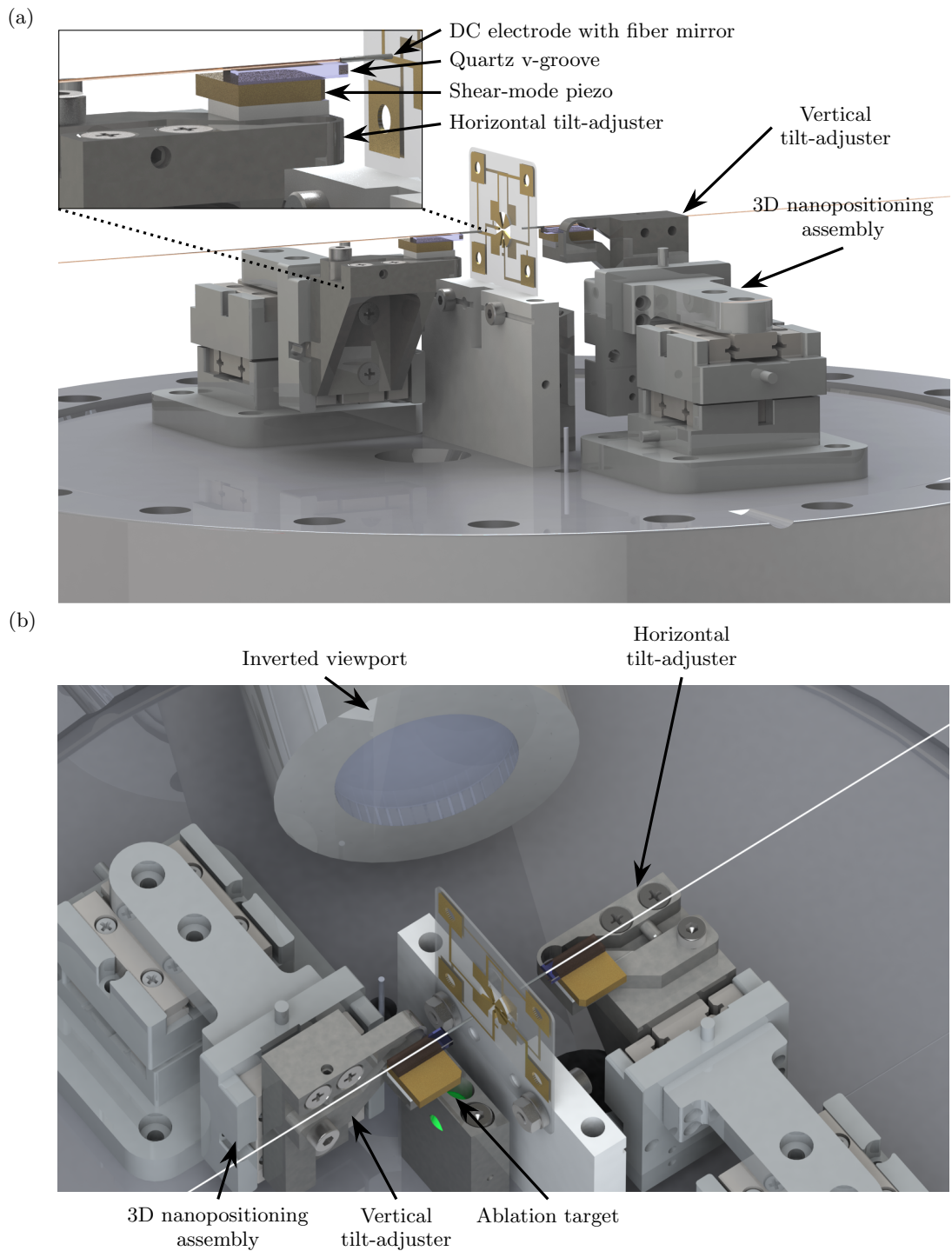

The ion-cavity system is located inside an ultra-high vacuum chamber at a pressure below $\SI{1E-10}{\milli\bar}$, the lowest pressure value that can be determined from the ion-pump current. The DC electrodes with integrated fiber mirrors are glued on quartz v-grooves~\footnote{Epoxy EPO-TEK 353ND}, each of which is glued on a shear-mode piezo~\footnote{Meggit PZ27}, as shown in Fig.~\ref{fig:figure2}. The length of the cavity is stabilized by applying a Pound--Drever--Hall feedback signal to one of the piezos~\cite{Black2001}. Each piezo is glued on a stainless-steel tilt-adjuster, with which the angle of each fiber mirror is aligned during the construction of the setup. Each tilt-adjuster is mounted on a 3D nanopositioning assembly~\footnote{The nanopositioning assembly consists of three SLC-1720-W-S
-UHVT-NM modules.}, which allows positioning of each fiber mirror along three axes over a range of $\SI{12}{\milli\meter}$ with a resolution of $\SI{1}{\nano\meter}$ for relative movements. In practice, we translate the fiber mirrors over a range of at most $\SI{1}{\milli\meter}$ along each axis.

For loading ions, we use single laser pulses with pulse energies around $\SI{150}{\micro\joule}$ at a wavelength of $\SI{515}{\nano\meter}$ to ablate neutral $^{40}$Ca atoms from a target, which is mounted $\SI{2}{\centi\meter}$ from the trap along the line of sight (Fig.~\ref{fig:figure2})~\cite{Hendricks2007}. 
Two objectives, each with a numerical aperture of $0.18$, are mounted inside inverted viewports. The objectives collect the ion fluorescence, which is guided to an electron-multiplying CCD camera and to a photomultiplier tube.

\section{\label{sec:Trap_simulation}Simulating the ion-cavity system}

We now consider two means by which the harmonic potential of the Paul trap may be distorted.
First, if surface charges are present on the fiber mirrors, their electric fields will shift the potential~\cite{Harlander2010,Ong2020probing}.
Second, if the nanopositioning assemblies are used to displace the DC electrodes and integrated fiber mirrors, the potential minimum may be displaced, or the confinement strength may change~\cite{Brandstaetter2013,mrquezseco2016novel}.
In this section, ion trap simulations are used to study both surface charges and electrode displacements. 
\subsection{Ion-trap potentials}\label{sec:concepts}
The trapping potential at a position $\mathbf{r} = (x,y,z)$ has three components: the pseudopotential $\phi_\mathrm{RF}$ generated by the RF electrodes, the potential $\phi_\mathrm{DC}$ generated by the DC electrodes, and the potential $\phi_\sigma$ due to any charges on the fiber facets~\cite{Ghosh1996}:
\begin{equation}
\phi_\mathrm{trap}(\mathbf{r}) = \phi_\mathrm{RF}(\mathbf{r}) + \phi_\mathrm{DC}(\mathbf{r}) + \phi_\sigma(\mathbf{r}) \mathrm{.}
\label{eq:1}
\end{equation}
To simulate the trapping potential, we follow the steps outlined in Ref.~\cite{Ong2020probing}.  We start by importing the geometry of the ion-cavity system as depicted in Fig.~\ref{fig:trap2} into finite-element analysis (FEA) software~\footnote{COMSOL Multiphysics 5.6}. We define our coordinate system to match the system indicated in Fig.~\ref{fig:trap2}, with the origin in the center of both the ion trap and the FFPC. The fiber-cavity length is set to $\SI{500}{\micro\meter}$ in this geometry.  Unless otherwise mentioned, the trap electrodes are grounded. 

We now determine the contributions $\phi_\mathrm{RF}$, $\phi_\mathrm{DC}$ and $\phi_\sigma$ separately for a $^{40}$Ca$^+$ ion (mass $m = 40\;\mathrm{u}$, charge $e$), starting with $\phi_\mathrm{RF}$.
In experiments, $\phi_\mathrm{RF}$ is generated by driving the wheel trap in one of two possible configurations. In the first configuration (RF-GND), we ground one pair of opposing RF electrodes and apply a driving signal with amplitude $V_\mathrm{RF}$ and frequency $\Omega_{\mathrm{rf}}$ to the other pair. In the second configuration (symmetric), we drive both RF electrode pairs such that the phase of the RF signal on one electrode pair is shifted by $\SI{180}{\degree}$ relative to the signal on the other pair.
To simulate the RF-GND configuration, we set a voltage $V_0 = \SI{1}{\volt}$ on one pair of RF electrodes.
To simulate the symmetric configuration, we set a voltage $V_0 = \SI{0.5}{\volt}$ on one pair of RF electrodes and a voltage $-V_0$ on the other pair. For both configurations, the FEA software simulates the electric field $\mathbf{E}(\mathbf{r})$, with which we calculate $\phi_\mathrm{RF}$ from the expression~\cite{house2008analytic}
\begin{equation}
\phi_\mathrm{RF}(\mathbf{r})=\frac{V_\mathrm{RF}}{V_0}\frac{e^2|\mathbf{E}(\mathbf{r})|^2}{4m\Omega_{\mathrm{rf}}^2} \mathrm{.}
\end{equation}

Next, we set a voltage $V_\mathrm{DC}$ on the DC electrodes and simulate $\phi_\mathrm{DC}(\mathbf{r})$.
Finally, to simulate $\phi_{\sigma}$, we set a homogeneous surface-charge density $\sigma$ on the facets of the fibers and assume the charges to be static. Since the fiber mirrors are located inside the DC electrodes, we do not consider charges on the sides of the fiber mirrors, in contrast to Ref.~\cite{Ong2020probing}. 

\begin{figure}

\begin{subfigure}{0.5\textwidth}
\caption{}\label{fig:figure3}
\includegraphics[scale=0.5]{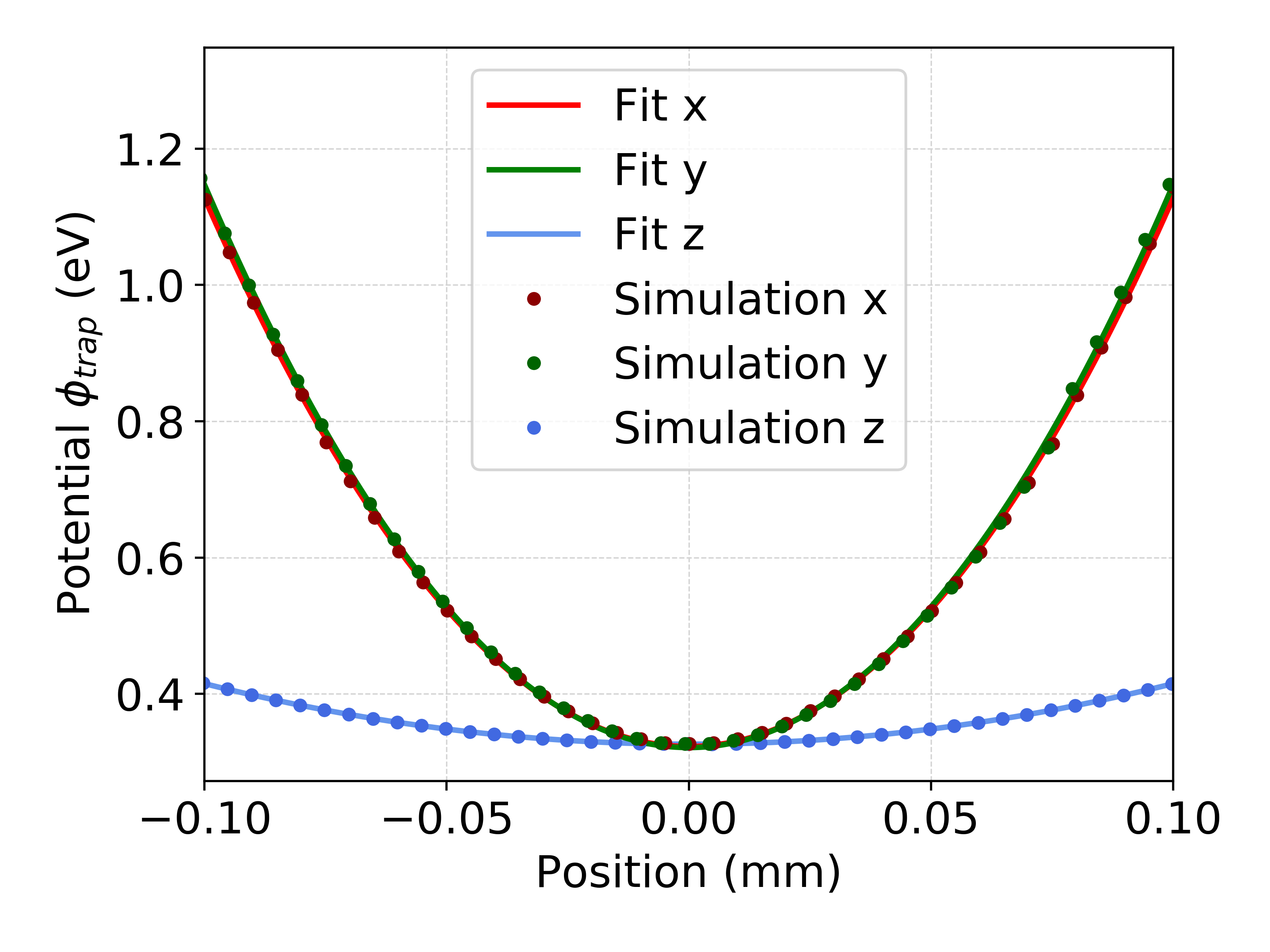}
\end{subfigure}
\begin{subfigure}{0.5\textwidth}
\caption{}\label{fig:figure3b}
\includegraphics[scale=0.5]{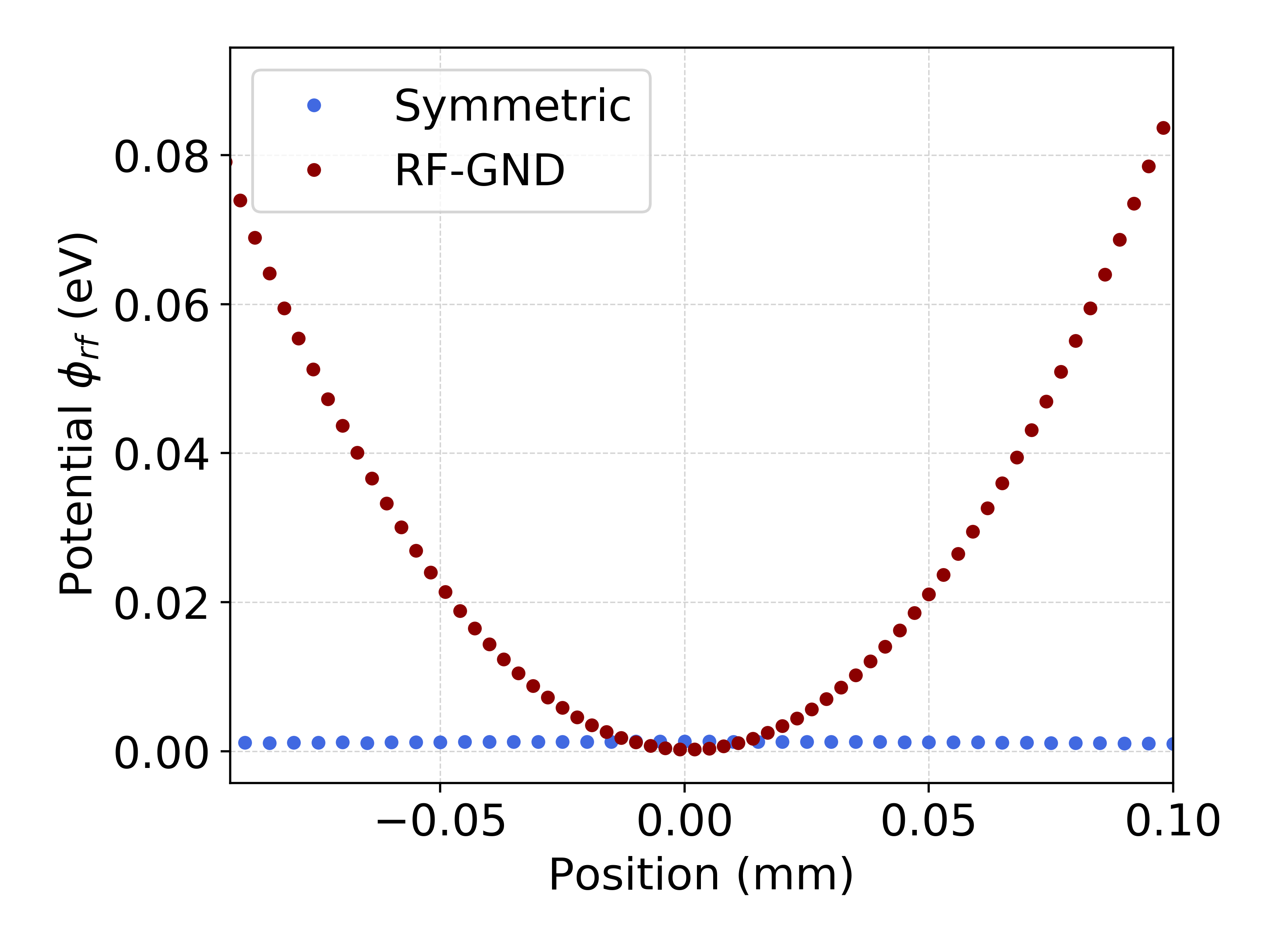}
\end{subfigure}
\caption{a) Simulated potential $\phi_\mathrm{trap}$ for the symmetric drive configuration with $V_\mathrm{RF} = \SI{160}{\volt}$ and $V_{\textrm{DC}} = \SI{1}{\volt}$, plotted for all three axes.  Fits of the data with Eq.~\ref{eq:harmonic} are also plotted.  b) Simulated RF potential $\phi_\mathrm{RF}$ for the symmetric and the RF-GND drive configurations, plotted for the z axis.}
\end{figure}

Following Eq.~\ref{eq:1}, we sum the three potentials to determine $\phi_\mathrm{trap}$ over the trapping region.
In Fig.~\ref{fig:figure3}, $\phi_\mathrm{trap}$ is plotted as a function of position along all three axes.
Here, the symmetric drive configuration is used with $V_\mathrm{RF}=\SI{160}{\volt}$, $V_\mathrm{DC} = \SI{1}{\volt}$, and no charges present on the fiber mirrors. 
We fit a harmonic oscillator potential \begin{equation}
	\phi(\mathrm{r})=\frac{1}{2}m\omega_r(r-r_0)^2+\phi_0 \label{eq:harmonic}
\end{equation} with offset $\phi_0$ to $\phi_\mathrm{trap}$ in order to extract the ion position $r_0 \in \{x_0,y_0,z_0\}$ and the trap frequency $\omega_{r}$ along an axis $r\in \{x,y,z\}$.
This fit yields $x_0=\SI{0(1)}{\micro\meter}$,  $y_0 = \SI{1(1)}{\micro\meter}$, $z_0=\SI{0(1)}{\micro\meter}$,
 $\omega_x = 2\pi\cdot {\SI{3.134(2)}{\mega\hertz}}$, $\omega_y = 2\pi \cdot \SI{3.174(2)}{\mega\hertz}$ and $\omega_z =2\pi\cdot \SI{1.041(1)}{\mega\hertz}$, where the uncertainties on the positions are set by the mesh resolution in the simulation.

In Fig.~\ref{fig:figure3b}, we compare $\phi_\mathrm{RF}$ along the z axis for the two drive configurations.
Over a range of $\SI{200}{\micro\meter}$ around the trap center,  the potential of the symmetric drive is constant at a value of approximately $\SI{1.22(8)}{\milli\electronvolt}$.  The potential of the RF-GND drive is harmonic with a minimum at $z=\SI{1(1)}{\micro\meter}$ and reaches a maximum value of $\SI{89.0(1)}{\milli\electronvolt}$. 
In the symmetric case, the $\SI{180}{\degree}$ phase shift of the two RF signals causes the electric field to vanish along the z axis~\cite{Paul1990}, while the asymmetric RF-GND configuration generates a vanishing field at only one point \cite{Berkeland1998}, meaning that ions displaced from this minimum will be subject to micromotion.
This statement holds true for all linear Paul traps, but for typical centimeter-scale trap lengths, the curvature of $\phi_\mathrm{RF}$ in the RF-GND configuration is negligible. However, for the $\SI{300}{\micro\meter}$-long wheel trap, the curvature of  $\phi_\mathrm{RF}$ becomes significant, as we will see in Sec.~\ref{sec:Comparison}.

\subsection{\label{sec:Surface_charges}Influence of surface charges}
We now include surface charges on the fiber facets in our simulations. The trapping potential is simulated for surface-charge densities $\sigma$ ranging from $0.1$ to $\SI{50}{\elementarycharge\per\micro\meter^2}$, with $V_\mathrm{RF}=\SI{160}{\volt}$ and $V_\mathrm{DC} = \SI{0}{\volt}$.  In each simulation, the same value of $\sigma$ is used for both fiber facets. In this first simulation of charge densities, values less than or equal to zero are not considered, since they lead to unstable trapping without a voltage on the DC electrodes. As in Sec.~\ref{sec:concepts}, we determine the trap frequencies, with uncertainties given by the standard deviations of the fit parameters.

\begin{figure}

\begin{subfigure}{0.5\textwidth}
\caption{}\label{fig:Charges}
\includegraphics[scale=0.5]{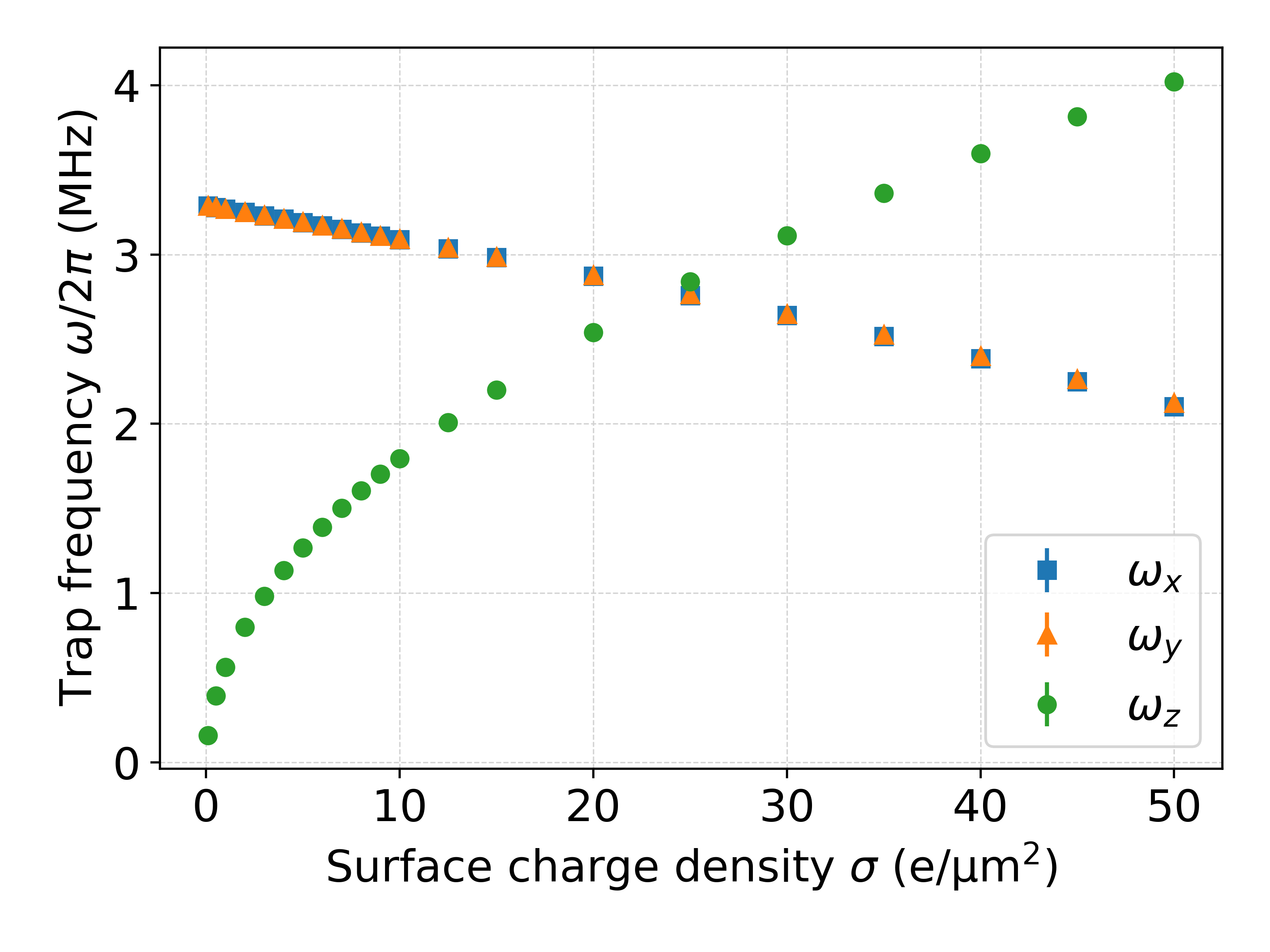}
\end{subfigure}
\begin{subfigure}{0.5\textwidth}
\caption{}\label{fig:Compensating}
\includegraphics[scale=0.5]{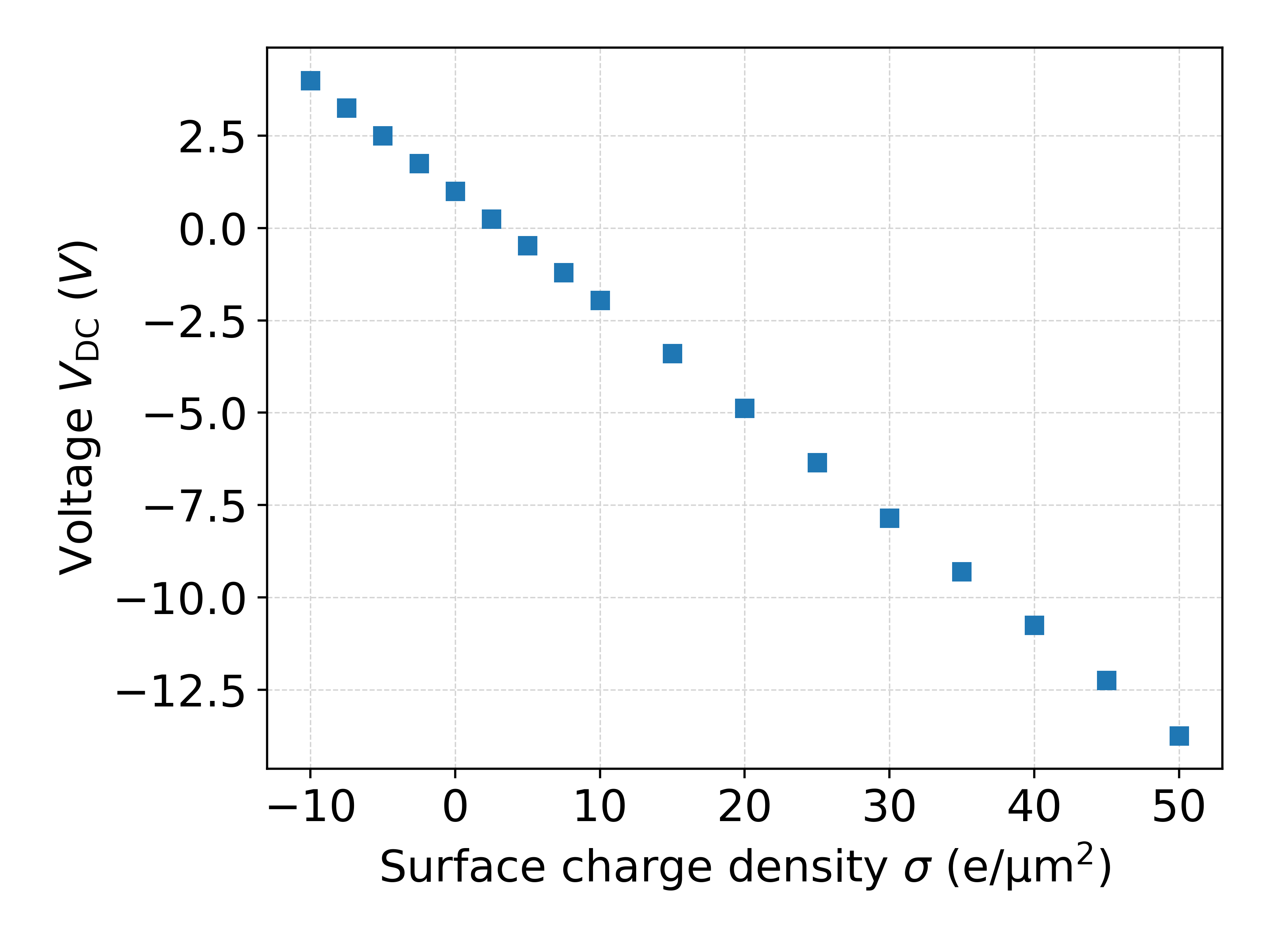}
\end{subfigure}
\caption{a) Motional frequencies $\omega_x$, 
	$\omega_y$, and $\omega_z$ plotted for surface-charge densities up to $\SI{50}{\elementarycharge\per\micro\meter^2}$ on the fiber mirrors. Error bars correspond to the standard deviation of the fit parameters and are too small to be visible. b) Voltage $V_\mathrm{DC}$ corresponding to an axial trap frequency $\omega_z = 2\pi\cdot\SI{1}{\mega\hertz}$ as a function of the surface-charge density.}
\end{figure}

Figure~\ref{fig:Charges} shows the trap frequencies along all three axes as a function of the surface-charge density. We observe that $\omega_z$ increases with increasing charge density while $\omega_x$ and $\omega_y$ decrease.  This is the same effect that one observes when increasing $V_\textrm{DC}$ for a linear Paul trap, since the surface charges and the applied voltage play the same role.  
Here we highlight another advantage of using the wheel trap for an integrated fiber cavity: charges on the fiber mirrors are equivalent to DC voltages on the endcaps.  In practice, it is difficult to add or remove surface charges in a controlled fashion \cite{Harlander2010,Ong2020probing}, but $V_\textrm{DC}$ provides a knob with which we can achieve the equivalent result.

As a proof of principle for this approach, 
we determine the value of $V_\mathrm{DC}$ that results in an axial trap frequency of $\omega_z =2\pi\cdot\SI{1}{\mega\hertz}$ for surface charge densities between $-10$ and $\SI{50}{\elementarycharge\per\micro\meter\squared}$. These values are plotted in Fig.~\ref{fig:Compensating}. This range of densities corresponds to the range from experiments reported in Ref.~\cite{Ong2020probing}. The compensation voltage decreases linearly with increasing surface charge density. Note that the approach works for any value of $\omega_z$; $\omega_z =2\pi\cdot\SI{1}{\mega\hertz}$ was simply chosen as a round number.

\subsection{Influence of the cavity position}\label{sec:fiber_position}

As a final consideration in our simulations, we vary the positions of the DC electrodes with integrated fiber mirrors in order to understand the effect on the trapping potential.  In the experimental setup, it is necessary to adjust the relative positions of the electrodes so that the mirrors form a cavity. The laser-ablation process results in mirror profiles that are centered with respect to the fiber facets with an uncertainty of $\SI{0.9}{\micro\meter}$~\cite{Ott2016a}.
Gluing the fibers into the DC electrodes results in a centering uncertainty of $\SI{30}{\micro\meter}$. Thus, we require a positioning range of $\SI{31}{\micro\meter}$. In addition, in order to position the cavity mode with respect to an ion, we will need to translate both mirrors and thus both electrodes.

\begin{figure}
\includegraphics[scale=0.5]{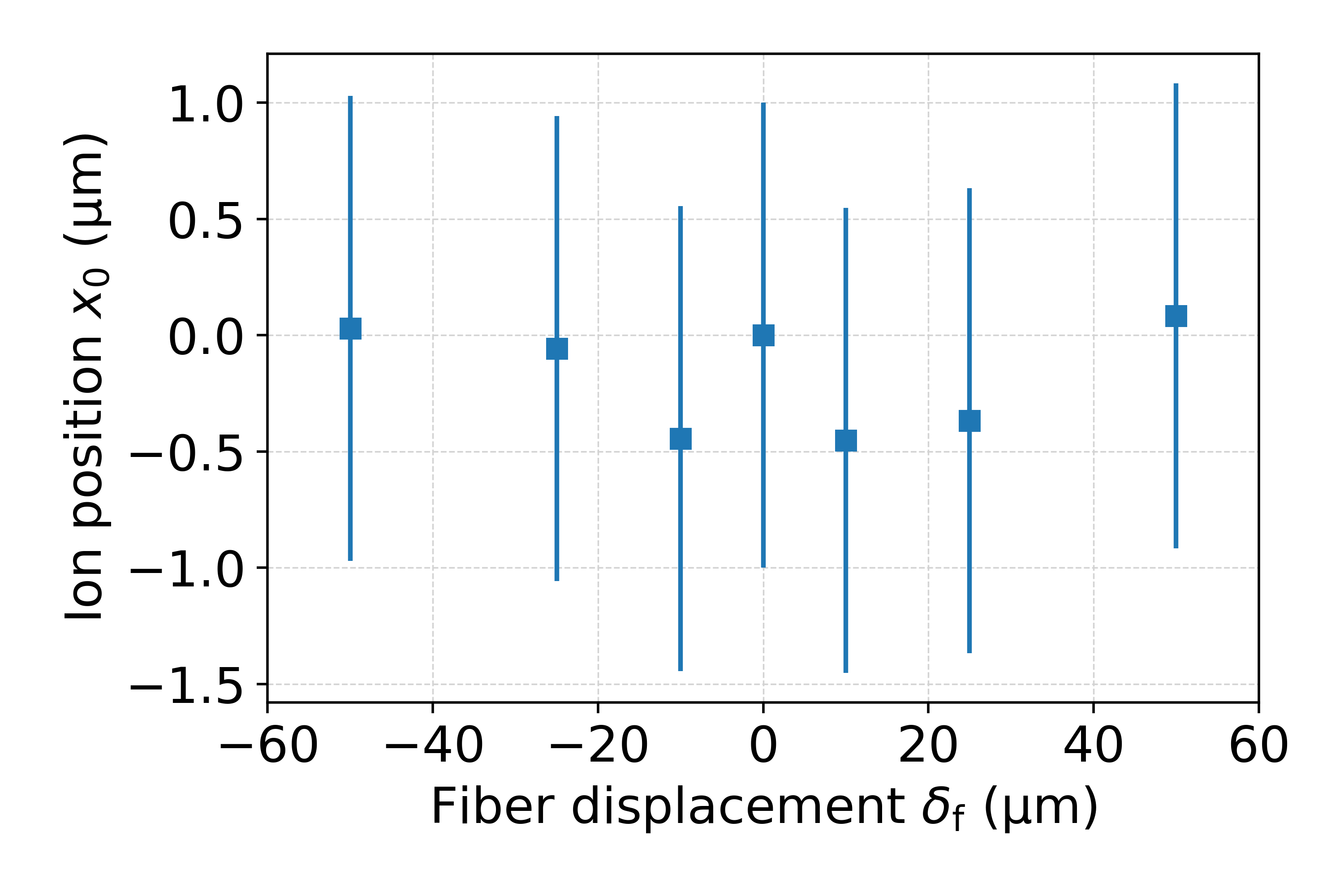}
\caption{Ion position $x_0$ as function of the position of the fiber mirrors along the x axis, indicated in Fig.~\ref{fig:trap2}.}\label{fig:Displacement}
\end{figure}

We displace both fiber mirrors along the x  axis for values  $\delta_f$  between $\SI{-50}{\micro\meter}$ and $\SI{50}{\micro\meter}$ and determine the position of the ion $\mathbf{r}_0 = (x_0,y_0,z_0)$, which is plotted in Fig.~\ref{fig:Displacement} for the x axis.
Again, we estimate a $\SI{1}{\micro\meter}$ uncertainty for the ion position from the resolution of the simulation mesh. 
Within the uncertainty, we observe no displacement of the ion, from which we conclude that it will be possible to position the fiber mirrors within our setup without affecting the ion position.


\section{Experimental tests}\label{sec:Measurements}


Prior to assembly of the ion-cavity system, we built an ion-trap test setup without integrated fiber mirrors. The DC electrodes described in Sec.~\ref{sec:Trap} are replaced by electrodes with
 $\SI{260(50)}{\micro\meter}$ inner diameter and $\SI{410(20)}{\micro\meter}$ outer diameter, separated by $\SI{3}{\milli\meter}$. We perform two experimental tests with this setup: first, a measurement of the micromotion, and second, measurements of the motional heating rates. The heating rates quantify how much electric-field noise couples from the environment to the motion of the ion~\cite{RevModPhys.87.1419}.

After assembling the ion-cavity system, we repeated the heating rate measurements. The earlier measurements without fiber mirrors allow us to distinguish between noise observed with the bare ion trap and noise due to the dielectric fiber mirrors, which are known sources of electric-field noise~\cite{Kumph_2016,Teller2021}. Subsequently, we tested the predictions of Sec.~\ref{sec:Surface_charges} regarding the influence of surface charges present on the fiber mirrors.

\subsection{Comparison of micromotion in RF-GND and symmetric configurations}\label{sec:Comparison}
 A symmetric configuration of the ion-trap drive leads to a vanishing electric field along the trap axis, as discussed in Sec.~\ref{sec:concepts}. Using the ion-trap test setup, we quantify the micromotion of a single $^{40}$Ca$^+$ ion for both symmetric and RF-GND configurations over a $\SI{20}{\micro\meter}$ range along the z axis.  We measure Rabi oscillations on the  $\ket{4^2\mathrm{S}_{1/2},m_j = +1/2}$ to $\ket{3^2\mathrm{D}_{5/2},m_j = +1/2}$ transition (qubit transition) and on the micromotion sideband of this transition. 
We determine the Rabi frequency $\Omega_Q$ of the qubit transition using a fitting model that assumes the Debye-Wallner coupling as the damping source~\cite{Wineland1998}. 
For the micromotion sideband, this approach is not applicable, since the period of Rabi oscillations is longer than the $\SI{259(11)}{\micro\second}$ coherence time of the qubit. We instead extract the Rabi frequency $\Omega_M$ by fitting a damped sinusoidal oscillation. Error bars correspond to one standard deviation of the fit parameters.


For $\Omega_M < \SI{1}{\kilo\hertz}$, we are unable to resolve multiple oscillations of the ion's state due to limitations in the experimental control hardware. For these Rabi frequencies, we estimate $\Omega_M$ from the first data point at which the excitation on the micromotion sideband overlaps with $0.5$, and the error of $\Omega_M$ is given by the quantum projection noise.
When displacing the ion along the z axis, the mean voltage on both electrodes is held constant in order to keep $\omega_\mathrm{z}$ constant. 
At each ion position, before determining $\Omega_Q$ and $\Omega_M$, we minimize the micromotion using the compensation electrodes. 

\begin{figure}
\includegraphics[scale=0.5]{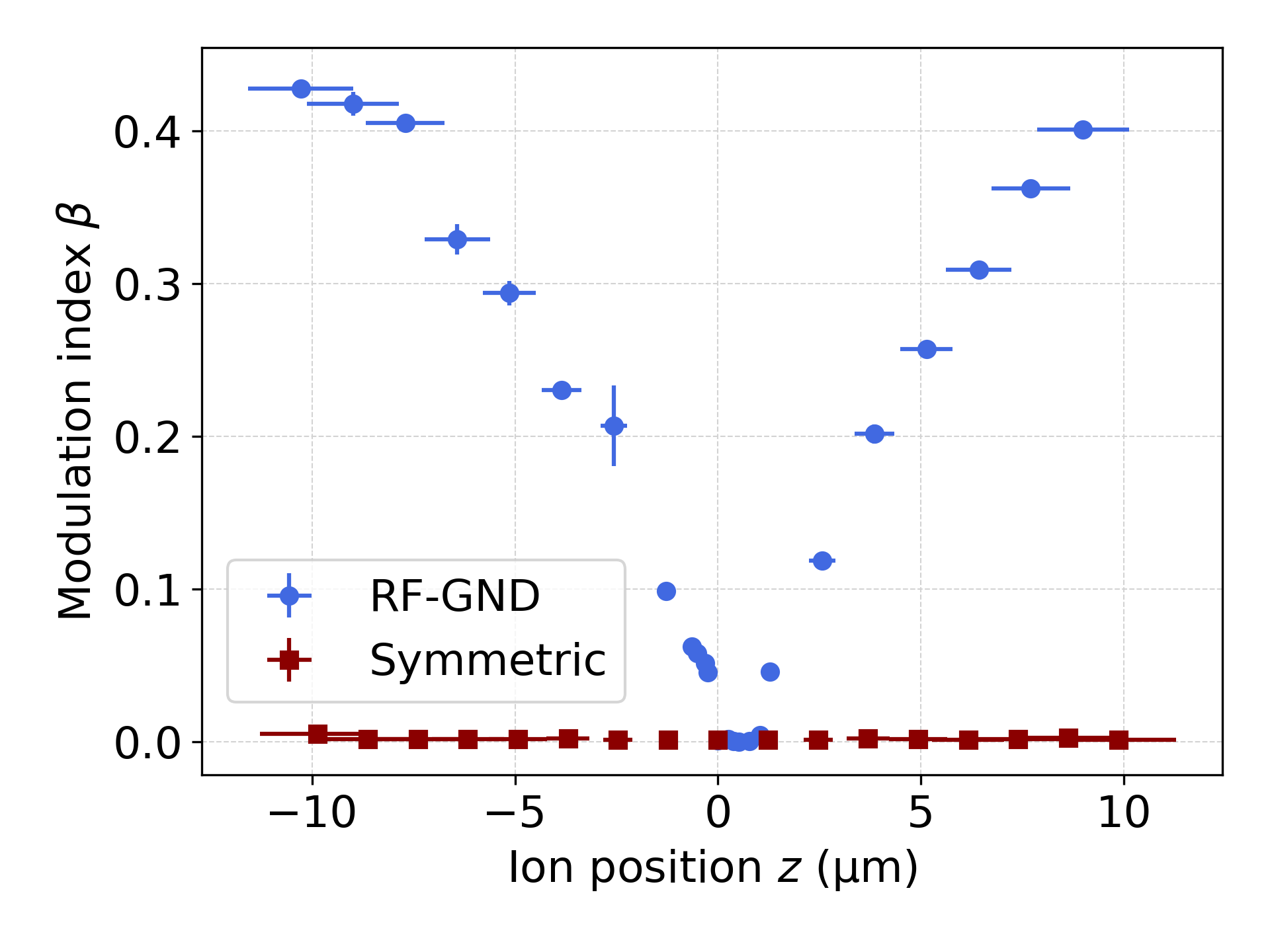}
\caption{Modulation index $\beta$ as function of the ion position for the RF-GND and symmetric drive configurations.}
\label{fig:Modulation_index}
\end{figure}

In Fig.~\ref{fig:Modulation_index}, the modulation index $\beta \approx 2\frac{\Omega_M}{\Omega_Q}$ is plotted as a function of the ion position.
The modulation index is proportional to the residual RF electric field and is thus a measure of micromotion~\cite{Berkeland1998}.  The micromotion vanishes at a single point for the RF-GND configuration, as expected from the simulations in Sec.~\ref{sec:concepts}.   In contrast, the micromotion vanishes over the full measurement range of $\SI{20}{\micro\meter}$ for the symmetric configuration.  The setup has been designed to couple multiple ions to the fiber-cavity mode. As a typical ion--ion distance is around $\SI{5}{\micro\meter}$, we see from Fig.~\ref{fig:Modulation_index}, that it is not possible to confine two ions in the RF-GND configuration without excess micromotion.  In contrast, in the symmetric configuration, it should be possible to confine at least five ions without excess micromotion. 

The ion position in Fig.~\ref{fig:Modulation_index} is calibrated as follows: we take an image of the ion at $z=\SI{0}{\micro\meter}$, displace the ion with the DC electrodes, and take an image of the displaced ion. From the two images, we calculate the ion displacement in units of pixels per volt. After measuring the micromotion, we load two ions into the trap and take a third image, from which we determine the distance $\delta z$ between the two ions in pixels.  We also calculate $\delta z$ in meters with the relation~\cite{James1998}
\begin{equation}
\delta z = \Big ( \frac{e^2}{2\pi \epsilon_0 m \omega_z^2} \Big )^{1/3} \mathrm{,}
\end{equation} 
thereby obtaining a conversion from pixels to meters.
Combining both conversions yields the ion displacement in units of meters per volt.
We obtain $\SI{1.28(16)}{\micro\meter\per\volt}$ for the RF-GND drive with $\omega_{z} = 2\pi \cdot \SI{1.517(1)}{\mega\hertz}$ and $\SI{1.23(18)}{\micro\meter\per\volt}$ for the symmetric drive with $\omega_{z} = 2\pi \cdot \SI{1.650(1)}{\mega\hertz}$. 

\subsection{\label{sec:Heating_rate}Heating rate measurements without fiber mirrors}
We measure the ion heating rate using sideband thermometry~\cite{Turchette2000}:
the ion is Doppler-cooled for $\SI{5}{\milli\second}$, followed by optical pumping and $\SI{10}{\milli\second}$ of sideband cooling. After a waiting time $t_\mathrm{w}$, we drive the red motional sideband of the qubit transition for $\SI{2}{\milli\second}$. This sequence is repeated for the blue sideband.  The phonon number is extracted from 100 of these measurements, and this process is then repeated 20 times for each waiting time in order to calculate the mean value and the sample standard deviation. For all measurements presented, we set the trap frequencies to $\omega_\mathrm{x}=2\pi\cdot\SI{3.399(1)}{\mega\hertz}$, $\omega_\mathrm{y} =2\pi\cdot \SI{3.229(1)}{\mega\hertz}$, and $\omega_\mathrm{z} =2\pi\cdot \SI{1.517(1)}{\mega\hertz}$.

\begin{figure}
\includegraphics[scale=0.5]{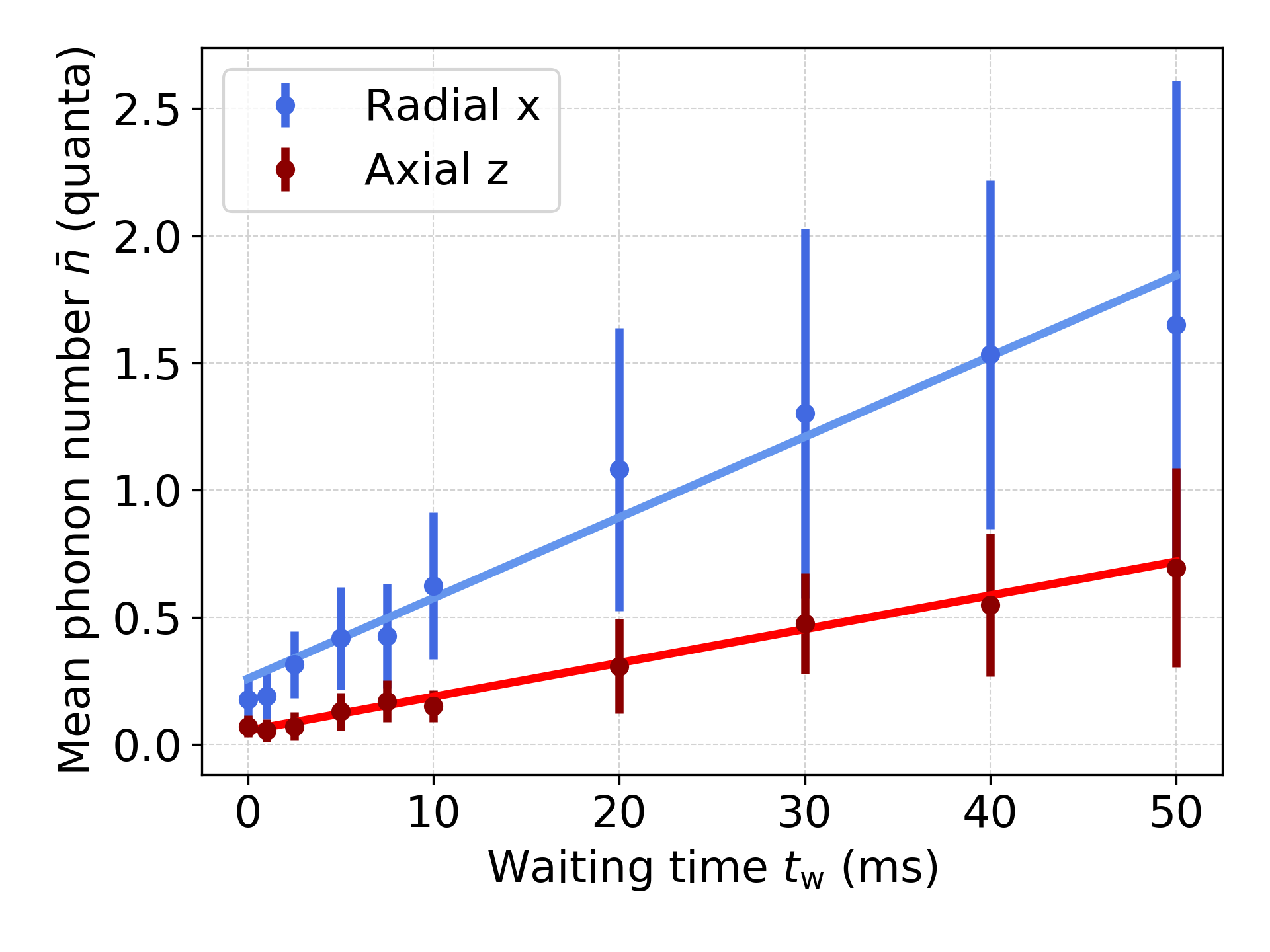}
\caption{Mean phonon number extracted via sideband thermometry as a function of the waiting time between sideband cooling and interrogation pulse, without fiber mirrors integrated in the setup. The solid lines represent weighted fits corresponding to heating rates of $\dot{\bar{n}}_z = \SI{13(3)}{\phonon\per\second}$ and $\dot{\bar{n}}_x=\SI{32(8)}{\phonon\per\second}$. Error bars represent the standard deviation calculated from 20 samples.}
\label{fig:Heating_rate}
\end{figure}

The phonon numbers of the axial mode and one radial mode are plotted in Fig.~\ref{fig:Heating_rate} for waiting times up to $\SI{50}{\milli\second}$.
From a weighted least-squares linear fit, we extract the heating rates $\dot{\bar{n}}_{x} = \SI{32(8)}{\phonon\per\second}$, $\dot{\bar{n}}_{y} = \SI{26(6)}{\phonon\per\second}$ and $\dot{\bar{n}}_{z} = \SI{13(3)}{\phonon\per\second}$. 
Similar rates have been measured with a $^{25}$Mg$^+$ ion in the original wheel trap~\cite{chen2017sympathetic,Brewer2019}.

\subsection{\label{sec:Heating_rate2}Heating rate measurements with fiber mirrors}
After integrating the fiber mirrors into the experimental setup, we measure the heating rates again. The distance between the fiber mirrors is set to $\SI{550}{\micro\meter}$ and the axial trap frequency to $2\pi\cdot\SI{1.636(5)}{\mega\hertz}$.   
We find that the red motional sideband is no longer suppressed by sideband cooling, so instead, the ion's temperature is determined from fits to Rabi oscillations~\cite{Roos,rowej2002transport,Haeffner2008,an2019distance,Teller2021}. 
The laser beam driving Rabi oscillations overlaps with all three motional modes, so we cannot determine the phonon number of each mode separately. As the contribution of the axial mode dominates by more than one order of magnitude, we express the mean phonon number $\bar{n}$ as a projection onto the z axis, as described in Ref.~\cite{Teller2021}. Note that $\bar{n}$ approximates the mean phonon number $\bar{n}_z$ of the axial mode, but $\bar{n}_z$ is smaller than $\bar{n}$.


\begin{figure}
\includegraphics[scale=0.5]{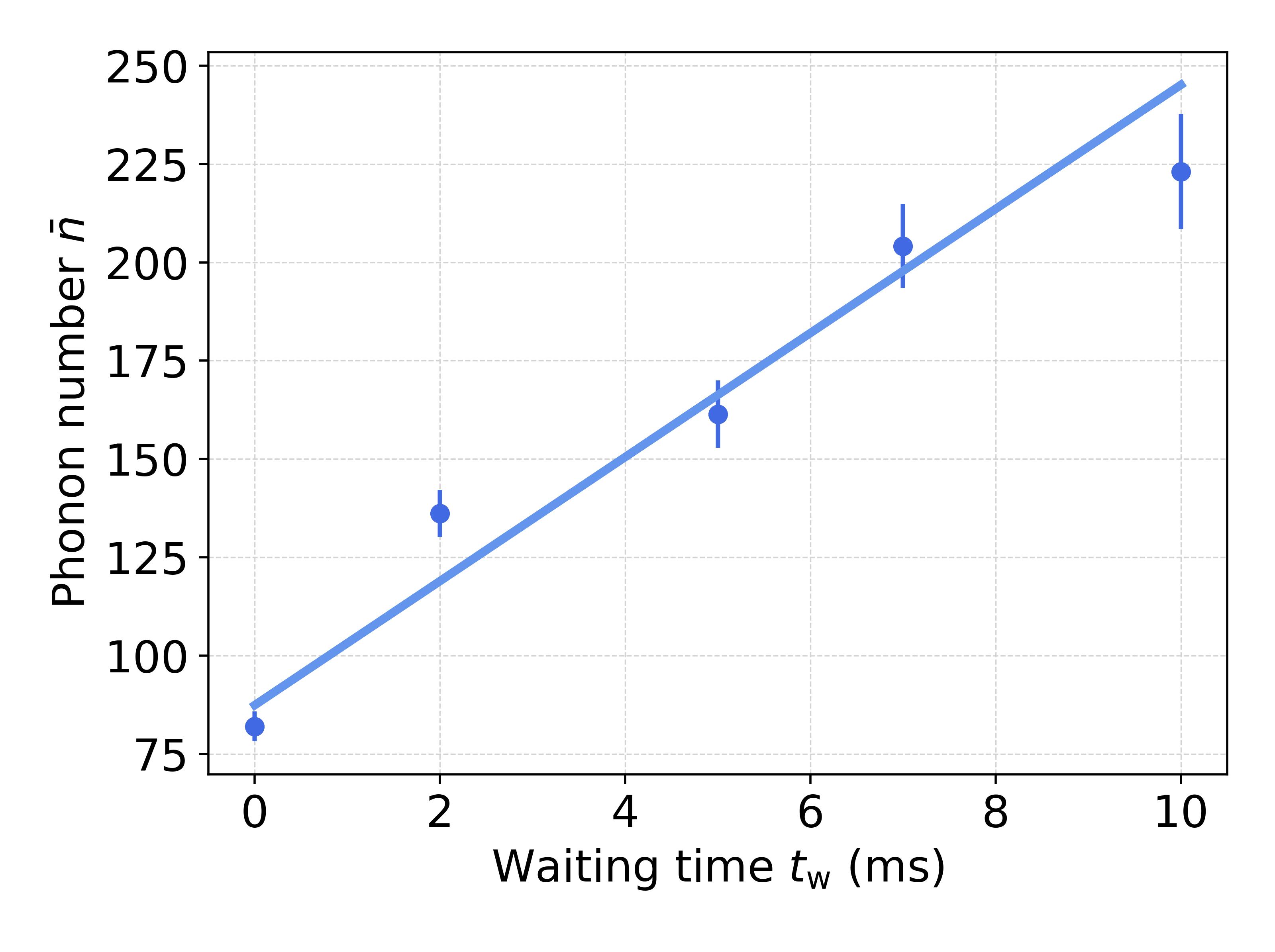}
\caption{Mean phonon number $\bar{n}$ as a function of the waiting time between Doppler cooling and interrogation pulse, with fiber mirrors integrated in the setup. The solid line represents a weighted fit corresponding to a heating rate of $\dot{\bar{n}} = \SI{14(2)}{\phonon\per\milli\second}$.}
\label{fig:Heating_rate2}
\end{figure}
In Fig.~\ref{fig:Heating_rate2}, $\bar{n}$ is plotted for a variable waiting time $t_\textrm{w}$ before the interrogation pulse.   
From a linear fit, we determine a heating rate of $\dot{\bar{n}} = \SI{14(2)}{\phonon\per\milli\second}$, three orders of magnitude larger than the rates in Sec.~\ref{sec:Heating_rate}. 
We attribute this higher rate to the presence of the fiber mirrors.  We have recently developed a model for ion heating based on dielectric losses in the fibers, which is supported by further experiments that we have conducted with this setup~\cite{Teller2021}. 

\subsection{Counteracting surface charges}
As a final test, we return to the predictions of Sec.~\ref{sec:Surface_charges} and adjust the trap electrode voltages to counteract the effects of surface charges on the fibers.
Starting from $L = \SI{507(8)}{\micro\meter}$, which is determined from a measurement of the cavity free-spectral range, we increase the length via the nanopositioning assemblies. 
For each value of $L$, we adjust the voltages on the DC electrodes such that the axial trap frequency is $\omega_z = 2\pi\cdot \SI{1.6(1)}{\mega\hertz}$ and the ion position remains fixed within $\SI{25}{\micro\meter}$.  Here,  $V_\mathrm{PC}$ and $V_\mathrm{MM}$ are the voltages on the DC electrodes containing the PC fiber and the MM fiber, respectively.
\begin{figure}
\includegraphics[scale=0.5]{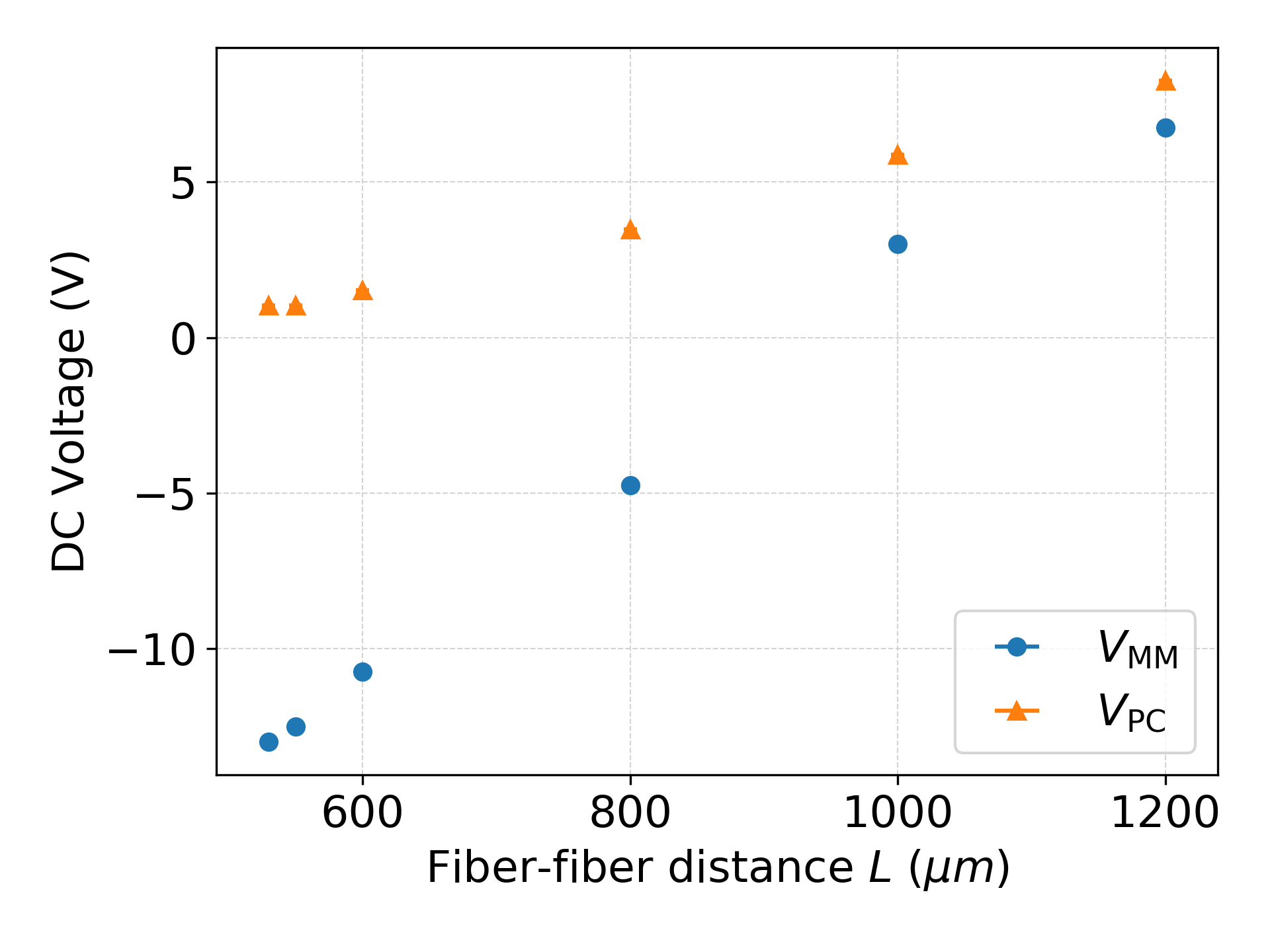}
\caption{Voltages $V_\mathrm{PC}$ and $V_\mathrm{MM}$ applied to the DC electrodes in order to keep both the ion position and trap frequency fixed over a range of fiber--fiber distances $L$. Error bars are too small to be visible.}
\label{fig:Pos_voltage}
\end{figure}
The uncertainties of $V_\mathrm{PC}$ and $V_\mathrm{MM}$ correspond to the precision of the voltage source. In Fig.~\ref{fig:Pos_voltage}, we plot $V_\mathrm{MM}$ and $V_\mathrm{PC}$ for fiber--fiber distances up to $\SI{1200}{\micro\meter}$. 
Both voltages increase for increasing distance, but $V_\mathrm{PC}$ is always positive, while $V_\mathrm{MM}$ takes on both negative and positive values over a range that is three times higher.
We attribute this difference to the presence of surface charges on the fiber facets.  This measurement shows that despite surface charges, a trapped ion can be confined at a fixed position over a range of cavity lengths.

\section{Conclusion}\label{sec:Conclusion}
We have designed and constructed an ion--cavity system with integrated fiber mirrors. The system is designed such that strong coupling of multiple ions to the fiber cavity will be possible without excess micromotion.  Simulations show that voltages on the DC electrodes compensate for surface charges on the fiber mirrors, and that translation of the fiber mirrors does not affect the ion position. 

Prior to the assembly of the system, we built an ion-trap test setup without the fiber mirrors and measured micromotion and heating rates, both of which are consistent with values in state-of-the-art ion traps for quantum information processing. After integration of the fiber mirrors, we observed a heating rate that was three orders of magnitude higher, which we attribute to the presence of the fibers. This observation led to a recent study on the role of dielectrics in ion traps~\cite{Teller2021}.

We have trapped an ion within the cavity for cavity lengths as short as $\SI{507(8)}{\micro\meter}$, and we have confirmed that voltages on the DC electrodes compensate for surface charges.  A next step will be to measure the coupling strength of single and multiple ions to the cavity field, followed by demonstrations of multi-ion protocols that have been previously implemented in macroscopic ion--cavity platforms, including quantum state-transfer~\cite{Casabone2015} and optimized collective coupling~\cite{Begley2016}.  Due to high coherent coupling rates and short cavity lifetimes, fiber-based systems may replace those bulkier setups, enabling scalable links between distributed ion-trap quantum computers.

All data presented and discussed in this article are available at Ref.~\cite{Zenodo}. The authors have no conflict of interest to disclose.

We thank David Leibrandt, Sam Brewer and Jwo-Sy Chen for discussions and insights on the design of the wheel trap. We thank Da An and Hartmut H\"{a}ffner for help with the design of the radio-frequency resonator used for the symmetric drive configuration of the ion trap.

This work was supported by the European Union's Horizon 2020 research and innovation program under Grant Agreement No. 820445 (Quantum Internet Alliance), by the U.S. Army Research Laboratory under Cooperative Agreement No. W911NF-15-2-0060, and by the Austrian Science Fund (FWF) under Project No. F 7109. P. H. acknowledges support from the European Union’s Horizon 2020 research and innovation programme under Grant Agreement No. 801285 (PIEDMONS). K.S. acknowledges support from the ESQ Discovery grant "Ion Trap Technology" of the Austrian Academy of Science. M.T. acknowledges support for the OptiTrap project under the Early Stage Funding Programme provided by the Vice-Rectorate for Research of the University of Innsbruck. 
\bibliography{cqed}

\end{document}